# Instrumentation and field tests to evaluate a rollover risk estimator for mobile machinery with mobile tools

**Lama Al Bassit [a,*], Bastien Laurent [a], Philippe Heritier [b] , Romain Duval [c] , Valérie Sulis [c] Roland Lenain [a]**

[a] INRAE, UR TSCF, Université Clermont Auvergne, 63178 Aubière, France

[b] INRAE, UR TSCF, Université Clermont Auvergne, 03150 Montoldre, France

[c] CETIM, 60300 Senlis, France

* Corresponding author. Email: lama.al-bassit@inrae.fr

### Abstract

Agricultural machines that carry mobile tools or implements have an internal dynamic that makes them more prone to rollover risk than machines with fixed configurations. Previous research has led to the development of algorithms capable of predicting rollover risk in real time by estimating relevant metrics, primarily the LTR (Load Transfer Ratio), with a focus on vehicles with static configurations. This paper presents the adaptation of these algorithms for machines with mobile tools and describes the field tests, conducted on a self-propelled sprayer, to evaluate their effectiveness in predicting rollover. The tests also aimed to acquire the necessary data to build an accurate numerical model of the vehicle and produce high-quality simulated tests. This paper describes the chosen instrumentation, the test procedure, and the experimental design of these tests. These field tests include static tests and dynamic tests. Static tests determine vehicle parameters, while dynamic tests measure variables related to the machine displacement and wheel/ground forces in real time. The initial results demonstrate the algorithms' ability to capture variations in LTR, which express lateral and longitudinal rollover risk. The data collected through these tests will enable the vehicle displacements to be replayed by simulation, producing simulated rollover accidents and evaluating the effectiveness of the developed algorithms in accident prevention.

## 1. Introduction

Rollover accidents involving mobile machinery are one of the main causes of occupational fatalities in the agricultural sector (Rondelli et al., 2018; Wang et al., 2024). Beyond formation to improve driver skills and safety procedure, technical solutions proposed to reduce rollover risk can be classified into two broad categories offering different levels of safety. The first comprises solutions designed to mitigate the consequences of a rollover accident, while the second comprises solutions designed to prevent them, offering a higher level of safety. The first category includes passive and active rollover protection structures (ROPS), which create a safety zone around machinery occupants and prevent crushing injuries in the event of a rollover (Franceschetti et al., 2019; Ojados-Gonzalez et al., 2016). It also includes systems that can detect rollovers and alert a remote rescue centre (Liu et al., 2013). The second category includes alarm systems that alert the driver to imminent rollover risk, enabling them to take appropriate action to prevent it (Hu et al., 2019). This category also includes systems that autonomously modify the vehicle's controlled variables (e.g. speed, steering and active suspension) to prevent rollover accidents (Richier, 2014; Zhu and Kan, 2022; Gao et al., 2022). For some technical solutions in both categories, especially the more effective ones in terms of safety, it is essential to estimate rollover imminence and predict hazardous situations.

To reduce the risk of agricultural machinery overturning, the INRAE TSCF research unit has been developing for several years real-time algorithms that estimate the rollover risk of outdoor machinery (Bouton et al., 2008; Richier et al., 2014). These algorithms have been tested in various conditions and on different types of vehicle without implements. The results demonstrated their effectiveness in simulations and field tests (Bouton et al., 2010; Lenain et al., 2012; Richier et al., 2015; Denis et al., 2017). These algorithms form the basis for autonomous active security devices.

The current project, conducted by the CETIM (Agricultural Machinery Technical Commission) and the INRAE TSCF unit, aims to enhance the stability and the safety of agricultural machinery carrying mobile implements. The internal dynamics of this type of machinery make them more prone to rollover risk than machinery with a static configuration.

The proposed approach to improving the stability of this type of agricultural vehicles involves developing and validating an online solution that can estimate rollover risk in order to prevent it. To this end, INRAE's stability risk assessment algorithms have been adapted for this type of machinery, and a numerical model of a vehicle with mobile implement has been developed. This enables simulated tests and accident risk scenarios to be conducted, thereby testing the risk estimation in various conditions. Field experiments are being conducted at the INRAE site in Montoldre using a self-propelled sprayer with a mobile ramp in order to test these new algorithms and collect data to calibrate the numerical model, which will enable highly accurate simulations.

This paper outlines the principles of the developed rollover risk estimation algorithms and describes the outdoor tests conducted on a self-propelled sprayer in order to collect the necessary data for validating the algorithms and calibrating the numerical model. It details the resources and instrumentation used, the static and dynamic tests carried out, and the results obtained.

## 2. Materials and Methods

The online stability estimation and rollover prevention approaches proposed in (Richier et al., 2011) and (Denis et al., 2017) are adapted here for off-road vehicles carrying mobile implements.

This section presents the theoretical modelling of this type of vehicle, which is used to compute the rollover metric. The metric adopted to represent stability is LTR (Load Transfer Ration), in both the lateral and longitudinal directions. The data required to evaluate the performance of the algorithms and the estimation process is identified. Static and dynamic tests are conducted on an instrumented Berthoud Raptor self-propelled sprayer to acquire this data. This section also details the technological solutions adapted for the instrumentation of the machinery, as well as the procedures for the tests conducted.

### 2.1. Theoretical model and stability metric calculation

The inertial parameters of a vehicle carrying a mobile implement vary depending on the implement's displacement relative to the vehicle body. Therefore, the position of the centre of gravity of the haul vehicle (G) and its inertia matrix are functions of the components of the vector connecting the centre of gravity of the vehicle's body (G1) to the centre of gravity of the implement (G2), (Figure 1). In this study, the implement's displacements are considered to be quasi-static (i.e. slow displacements with a velocity approaching zero), and it is assumed that the vehicle's inertia matrix remains diagonal in the vehicle frame (G; $X_v$,$Y_v$,$Z_v$) for all implement positions.

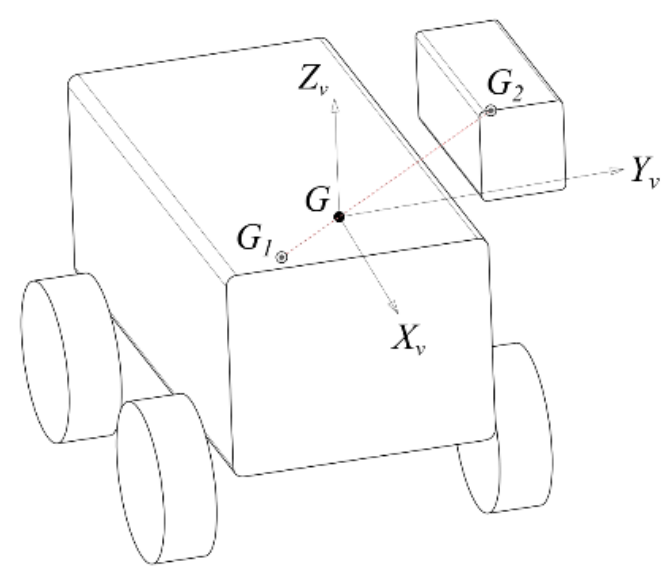

*Figure 1. Vehicle with moving implement*

The calculation of online lateral and longitudinal load transfer is achieved through the establishment of a dynamic model of the vehicle, based on three 2D projections: yaw, roll and pitch. As illustrated in Figure 2, the fundamental principle of dynamics is applied to each of these three models. The adopted parameters and variables are listed below:

- $m$ is the vehicle mass, including implement.
- $\alpha_r$ and $\alpha_p$ are the bank angles of the terrain in the roll and pitch projections.

- $L_f$ and $L_r$ are the front and the rear vehicle half-wheelbase lengths; they vary depending on the implement's position.
- $T_l$ and $T_r$ are the left and the right vehicle half-track lengths; they vary depending to the implement's position.
- $H$ is the distance between the roll centre, or the pitch centre, and the vehicle centre of gravity (G); it varies depending on the implement's position
- $I_x$, $I_y$, and $I_z$ are the roll, pitch and yaw moments of inertia; these also vary depending to the implement's position.
- $\psi$ is the vehicle yaw angle.
- $\varphi_r$ and $\varphi_p$ are the roll and the pitch angles of the suspended mass associated with roll and pitch dynamics.
- $\delta_f$ is the front steering angle; the rear steering angle is considered null.
- $v$ is the linear velocity at the centre of the rear axle.
- $u$ is the linear velocity at the roll centre O'.
- $\beta$ is the sideslip angle at the roll centre O'.
- $\beta_f$ and $\beta_r$ are the sideslip angles at the front and rear axles.
- $P = mg$ is the gravitational force acting on the suspended masse m, with g being the gravitational acceleration.
- $F_{Le}$ and $F_{Ri}$ are the normal components of the tyre/ground contact forces applied to the left and right sides of the vehicle.
- $F_{Fr}$ and $F_{Re}$ are the normal components of the tyre/ground contact forces applied to the front and rear of the vehicle.
- $F_f$ and $F_r$ are the lateral forces applied to the front and rear of the vehicle, respectively. A linear model of tire/ground contact is adopted here:

  $$F_f = C_f\,\beta_f \quad \text{and} \quad F_r = C_r\,\beta_r$$

  where the cornering stiffnesses, $C_f$ and $C_r$, are considered variable.
- $F_{ar}$ and $F_{ap}$ are roll and pitch restoring forces involving roll stiffness ($k_r$) and damping ($b_r$) coefficients and pitch stiffness ($k_p$) and damping ($b_p$) coefficients:

  $$\overrightarrow{F_{ar}} = \frac{1}{H}(k_r\varphi_r + b_r\dot{\varphi}_r)\overrightarrow{Y_v} \quad \text{and} \quad \overrightarrow{F_{ap}} = \frac{1}{H}(k_p\varphi_p + b_p\dot{\varphi}_p)\overrightarrow{X_v}$$

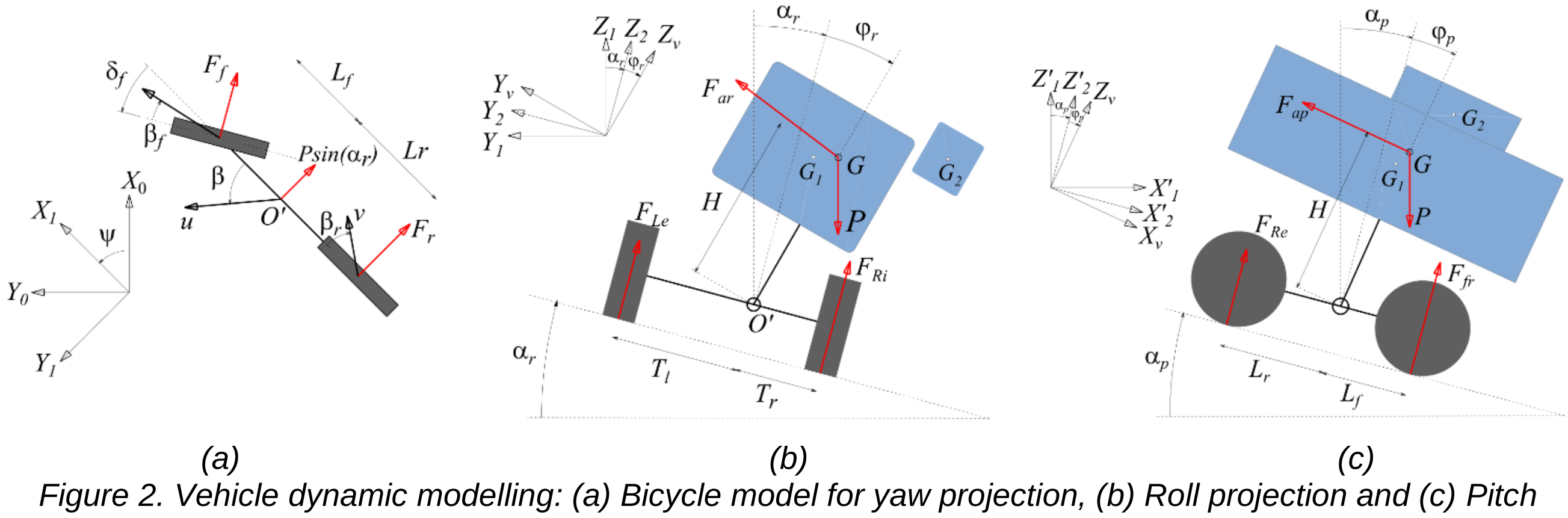

*Figure 2. Vehicle dynamic modelling: (a) Bicycle model for yaw projection, (b) Roll projection and (c) Pitch projection*

The yaw 2D projection (Figure 2(a)) is based on a simplified bicycle vehicle model. Applying the fundamental principle of dynamics yields the motion equations in the yaw frame. Assuming the angles are small and neglecting the non-influential variables (Richier, 2014), these equations can be written as follows:

$$\ddot{\psi} = \frac{-C_f L_F^2 \cos\delta_F - C_r L_R^2}{u I_z \cos\alpha_r}\dot{\psi} + \frac{-C_f L_F \cos\delta_F + C_r L_R}{I_z \cos\alpha_r}\beta + \frac{C_f L_F \cos\delta_F}{I_z \cos\alpha_r}\delta_F$$

$$\dot{\beta} = \left(\frac{-C_f L_F \cos(\delta_F - \beta) + C_r L_R \cos\beta}{u^2 m} - \cos\alpha_r\right)\dot{\psi} + \frac{-C_f \cos(\delta_F - \beta) - C_r \cos\beta}{um}\beta + \frac{C_f \cos(\delta_F - \beta)}{um}\delta_F - \frac{g}{u}\cos\beta \sin\alpha_r$$

This yaw model is used in an observer to estimate the non-measured variables required for calculating the lateral stability metric. The observer's algorithms are the same as those detailed in Richier et al. (2011). Figure 3 shows an overview of the observer principle. The measured variables that constitute the observer's input are velocity ($v$), steering angle ($\delta_f$), yaw rate ($\dot{\psi}$) and bank angle ($\alpha_r$) of the terrain. The estimated variables are sideslip angle ($\beta$), global lateral force ($F=F_f + F_r$) and cornering stiffness ($C$), where the front and rear cornering stiffnesse are considered equal ($C_f = C_r = C$).

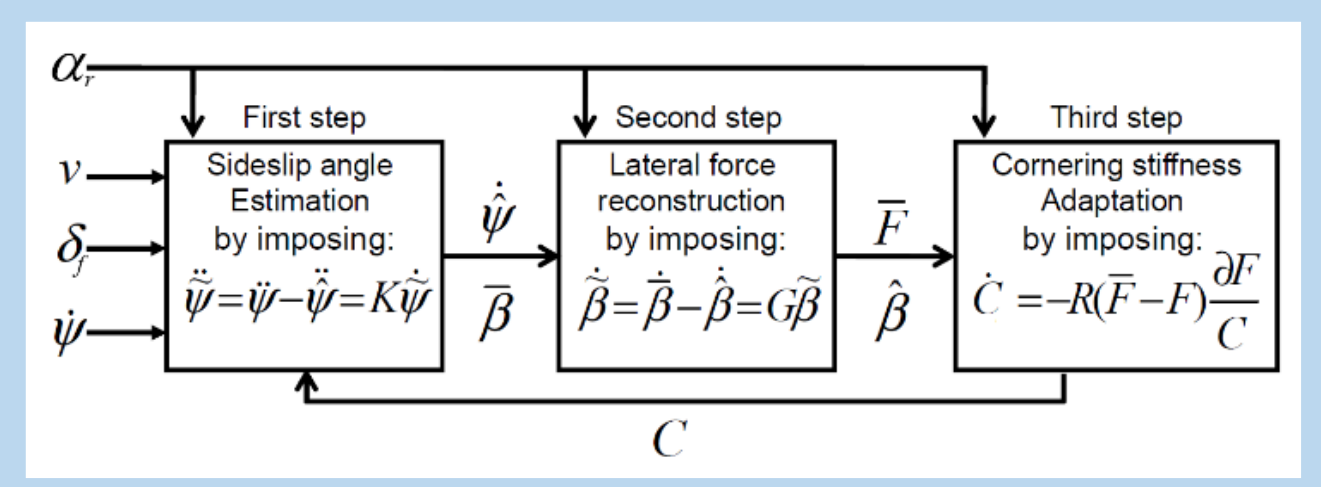

*Figure 3. Observer overview (Richier et al., 2011), in which the gains K and G are strictly negative, and R is strictly positive. The symbols bar, hat and tilde are used to designate control variables (inputs), estimated variables and error variables, respectively*

The roll model depicted in Figure 2(b) allows the dynamics of roll angle $\varphi_r$ and the normal forces ($F_{Le}$ *and* $F_{Ri}$) to be expressed as functions of $\varphi_r$ and other measured or estimated variables. Assuming the sideslip angle is small ($\sin(\beta)=\beta$) and neglecting the derivatives of the bank angle of the terrain, these equations can be written as follows:

$$\ddot{\varphi}_r = \frac{1}{H\cos\varphi_r}\left(H\dot{\varphi}_r^{\,2}\sin\varphi_r + H\dot{\psi}^2\sin(\varphi_r + \alpha_r)\cos\alpha_r + u\dot{\psi}\cos\beta\cos\alpha_r + \dot{u}\beta + u\dot{\beta} - \frac{F_{ar}}{m}\cos\varphi_r + g\sin\alpha_r\right)$$

$$F_{Le} + F_{Ri} = m\left(-H\ddot{\varphi}_r\sin\varphi_r - H\dot{\varphi}_r^{\,2}\cos\varphi_r + g\,\cos\alpha_r - \frac{F_{ar}}{m}\sin\varphi_r - H\dot{\psi}^2\sin(\varphi_r + \alpha_r)\sin\alpha_r - u\dot{\psi}\sin\alpha_r\right)$$

$$F_{Le} - F_{Ri} = \frac{2}{(T_l + T_r)}\left(I_x\ddot{\varphi}_r + \dot{\psi}^2(I_z - I_y)\sin(\varphi_r + \alpha_r)\cos(\varphi_r + \alpha_r) - \left(H\sin\varphi_r + \frac{T_l - T_r}{2}\right)(F_{Le} + F_{Ri})\right)$$

The pitch model depicted in Figure 2(c) allows the dynamics of pitch angle $\varphi_p$ and the normal forces ($F_{Fr}$ *and* $F_{Re}$) to be expressed as functions of $\varphi_p$ and other measured or estimated variables. Assuming the same hypothesis as before, these equations can be written as follows:

$$\ddot{\varphi}_p = \frac{1}{H\cos\varphi_p}\left(H\dot{\varphi}_p^{\,2}\sin\varphi_p + H\dot{\psi}^2\sin(\varphi_p + \alpha_p)\cos\alpha_p + u\dot{\psi}\sin\beta\cos\alpha_p - \dot{u} + u\dot{\beta}\beta + \frac{F_{ap}}{m}\cos\varphi_p + g\sin\alpha_p\right)$$

$$F_{Re} + F_{Fr} = m\left(-H\ddot{\varphi}_p\sin\varphi_p - H\dot{\varphi}_p^{\,2}\cos\varphi_p + g\,\cos\alpha_p + \frac{F_{ap}}{m}\sin\varphi_p - H\dot{\psi}^2\sin(\varphi_p + \alpha_p)\sin\alpha_p - u\dot{\psi}\beta\sin\alpha_p\right)$$

$$F_{Re} - F_{Fr} = \frac{2}{(L_f + L_r)}\left(I_y\ddot{\varphi}_p + \dot{\psi}^2(I_z - I_x)\sin(\varphi_p + \alpha_p)\cos(\varphi_p + \alpha_p) - \left(H\sin\varphi_p + \frac{L_r - L_f}{2}\right)(F_{Re} + F_{Fr})\right)$$

Lateral and longitudinal load transfer ratios ($LTR_{La}$ and $LTR_{Lo}$) are chosen as the relevant stability criteria for lateral and longitudinal stability, respectively. They are mathematically defined by the following equations:

$$LTR_{La} = \frac{F_{Le} - F_{Ri}}{F_{Le} + F_{Ri}}, \qquad LTR_{Lo} = \frac{F_{Fr} - F_{Re}}{F_{Re} + F_{Fr}}$$

LTR values vary between -1 and +1. These extreme values correspond to a vehicle with one side lifted off the ground. The closer a vehicle's LTR approaches +/-1, the more hazardous its situation becomes. Conversely, the safer the vehicle's situation, the closer the LTR is to 0.

The roll model equations allow the $LTR_{La}$ to be calculated. The variable $\varphi_r$ is obtained by integrating the equation that gives its second derivative. The other required variables are either measured or estimated by the observer. The pitch model equations allow the $LTR_{Lo}$ to be calculated. Here, the hypothesis of a steady-state pitch is considered, meaning that the time derivatives of $\varphi_p$ are nullified.

The sideslip angle $\beta$, which has no influence on longitudinal rollover, is considered null. The variable $\varphi_p$ and the $LTR_{Lo}$ are then obtained as functions of the measured variables only.

### 2.2. Data and process for stability estimation

The data required for an online LTR calculation can be divided into two categories: the vehicle's own parameters and its movement variables.

The vehicle's own parameters include geometric parameters (track and wheelbase), inertial parameters (mass, inertia matrix and position of the centre of gravity in a fixed vehicle frame), and restoring parameters (roll and pitch stiffness and damping coefficients). Without a moving implement or internal dynamics, the values of these parameters could be considered fixed. However, for vehicles with a mobile implement, certain of these parameters, notably the centre of gravity position and the inertia matrix are variable. Their values depend on the implement's position.

The vehicle's movement variables are related to the vehicle's control variables (velocity, acceleration, steering angle and yaw ratio) and the terrain conditions (bank angles in the roll and pitch projections).

Figure 4 summarises the stability estimation process. Implement displacement are measured to enable the vehicle's parameters to be calculated online. These parameters, along with the other measured values, feed the stability estimation algorithms.

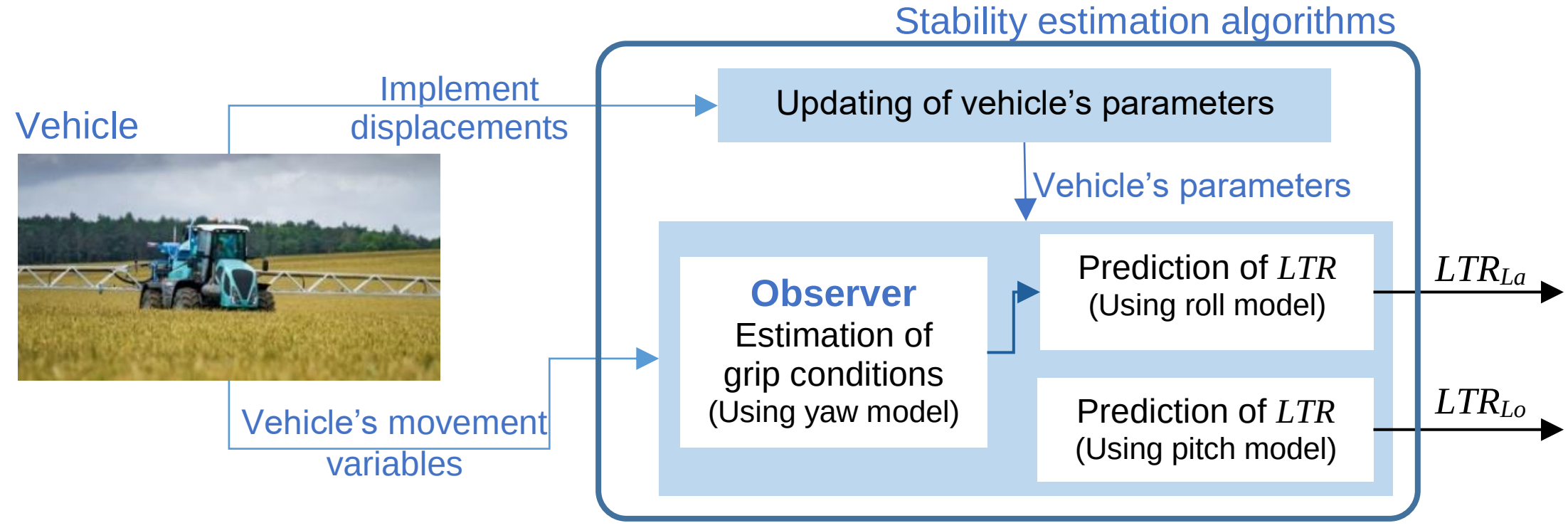

*Figure 4. Stability estimation process*

### 2.3. Machinery instrumentation for data collection

In order to collect data for evaluating the stability algorithms and creating an accurate digital model for future simulation tests, a Berthoud Raptor self-propelled sprayer with a tank capacity of approximately six tons (see Figure 4) is fitted with sensors. This machinery has a boom (a ramp) with nozzle holders that can cover a large working width (36 m) and can be adjusted to different heights. This mobile boom is referred to as an "implement" here.

The sensors and other on-board equipment used to equip this machine are detailed below:

#### *2.3.1. Sensors and equipment for load transfer estimation*

The implement (the ramp) is attached to the vehicle body via a parallelogram mechanism. Its displacement occurs in a plane. Wire distance sensors are used to measure the output of one of the implement cylinders (see Figure 5 (a)). The relationship between the implement cylinder output and the position of the vehicle centre of gravity is determined through static trials (see section 2.4)

An inertial module unit (IMU), which integrates a GPS, is located at the rear axle of the vehicle. This allows the velocity, acceleration, yaw rate and inclination angles of the vehicle to be measured.

To obtain the steering angle, a wire distance sensor is fixed to one of the front steering cylinders (Figure 5(b)). The relationship between the cylinder output and the steering angle of the wheel is calibrated using a highly accurate laser tracker. (Figure 5(c)).

To estimate load transfer online, a display module and an electronic control unit containing stability estimation algorithms are installed in the driver's cab of the vehicle.

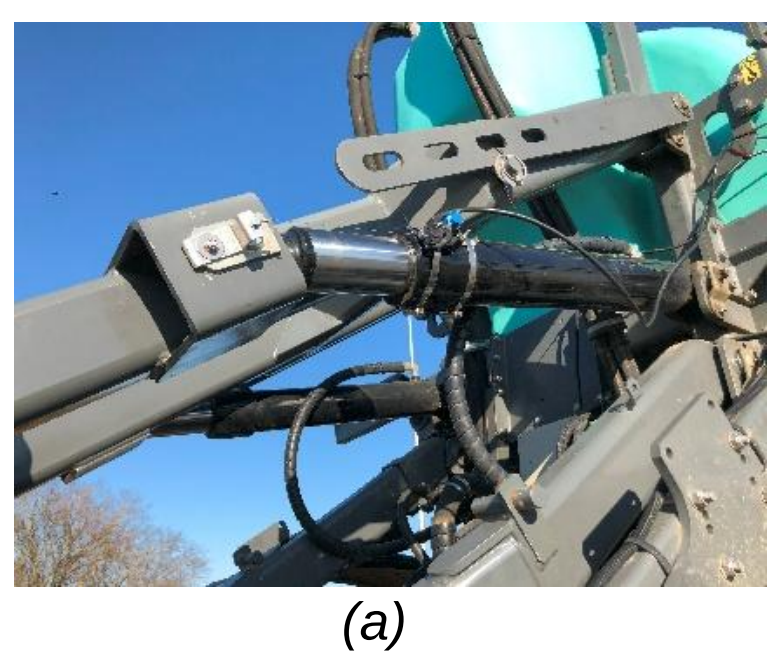
(a)

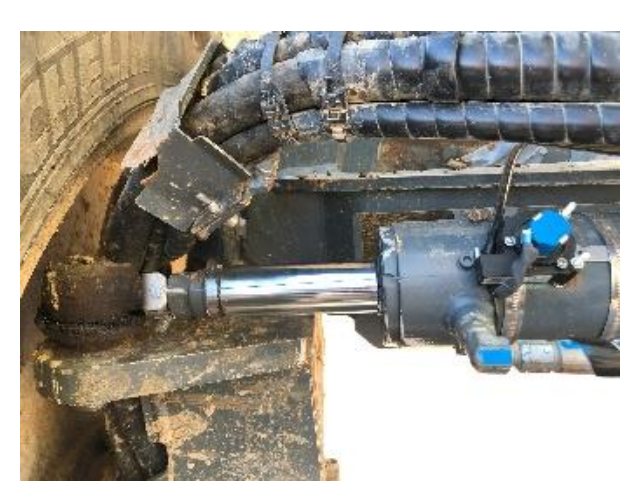
(b)

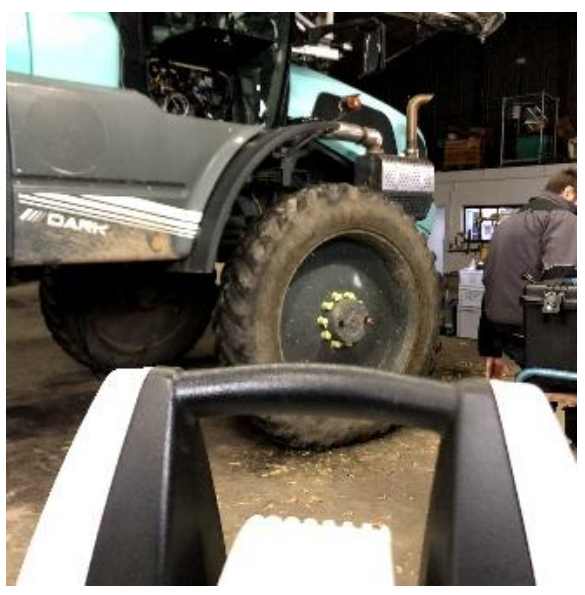
(c)

*Figure 5.Wire sensors fitted to (a)an implement cylinder and (b) a steering cylinder, (c) Laser tracker to calibrate steering angle*

*2.3.2. Sensors for load transfer measurement*

To measure online load transfer and compare it to the estimated value, wheel/soil force measurements are required. After reviewing possible solutions for measuring this type of force, the project team opted for a measurement solution based on strain gauges. The Berthoud Raptor machinery features an adjustable track, ensured by translation joints activated by cylinders. Each joint comprises two square profiles that fit into each other; one is fixed to the chassis and the other to the wheel assembly. When the track reaches its maximum value, the square profiles undergo bending, which is measured in order to deduce the forces applied to the wheels and, consequently, the LTR. Four square profiles, one for each wheel, are fitted with strain gauges. Each beam has sixteen gauges that measure bending efforts in two sections of the beam.

Calibration tests were carried out to determine the relationship between the measured strains and the vertical and longitudinal forces applied to the wheels. Weighing scales and a weighbridge were used to measure the forces between each wheel and the ground for different levels of vehicle tank filling. For longitudinal forces, calibration was achieved by applying forces to the chassis introduced by other machinery attempting to tow the vehicle. The traction forces between the two vehicles were measured using a load cell.

*2.3.3. Equipements and sensors for additional data*

To enable more accurate localisation of the vehicle during dynamic testing, a GPS RTK has been added to the aforementioned sensors.

To provide a detailed characterisation of the vehicle, pressure and wire distance sensors are used to measure the pressure and height variation of the four air suspension bellows (see Figure 6(a)). Cameras are also used to collect contextual data for each dynamic test.

An on-board acquisition centre (Figure 6(b)) receives data from all the sensors.

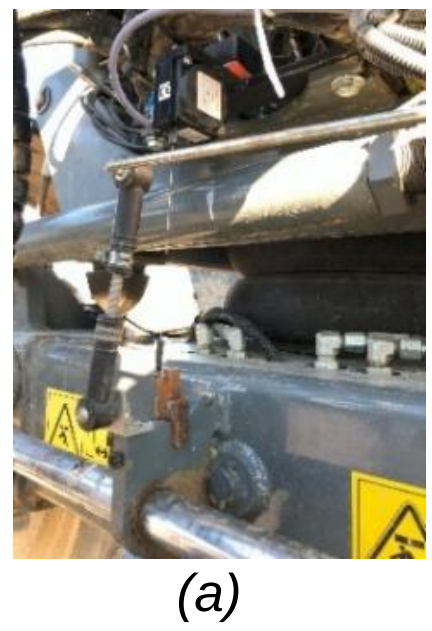
(a)

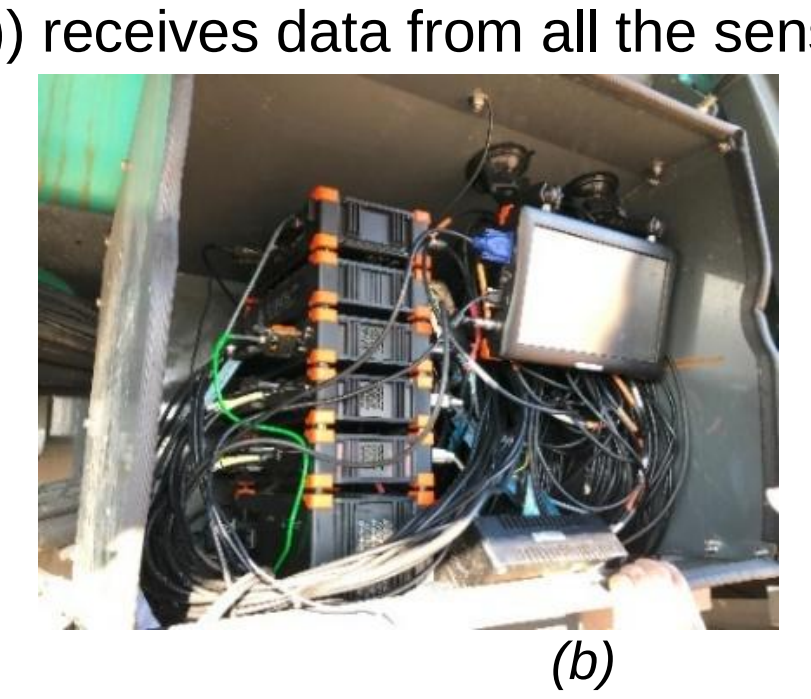
(b)

*Figure 6. (a) Wire sensor measuring height variations in a suspension bellows and (b) On-board acquisition centre*

**2.4. Static and dynamic tests**

Static and dynamic tests are conducted at the INRAE site in Montoldre, in the Allier department (03). The measurement equipment available on site, as well as the test fields and track of the AgroTechnoPôle[1] are made available to the project team to carry out these tests.

*2.4.1. Static tests*

Static tests aim to determine the vehicle's inertial and restoring parameters, as well as how these vary in relation to the configuration of the implement (folded or unfolded), its displacements, and the filling of the vehicle tank.

Different-height wedges are used to adjust the inclination of the vehicle (see Figure 7), while weight scales measure the distribution of wheel-ground forces at each inclination. These measurements are then used, alongside mathematical relations, to deduce the position of the centre of gravity and the roll and pitch rigidities of the vehicle.

All trials are repeated with the implement both folded and unfolded and with the tank at three different levels of filling: empty (total mass = 11,8 tons), half-filled (total mass = 14,1 tons) and full (total mass = 17 tons). This provides six sets of parameters for the machinery.

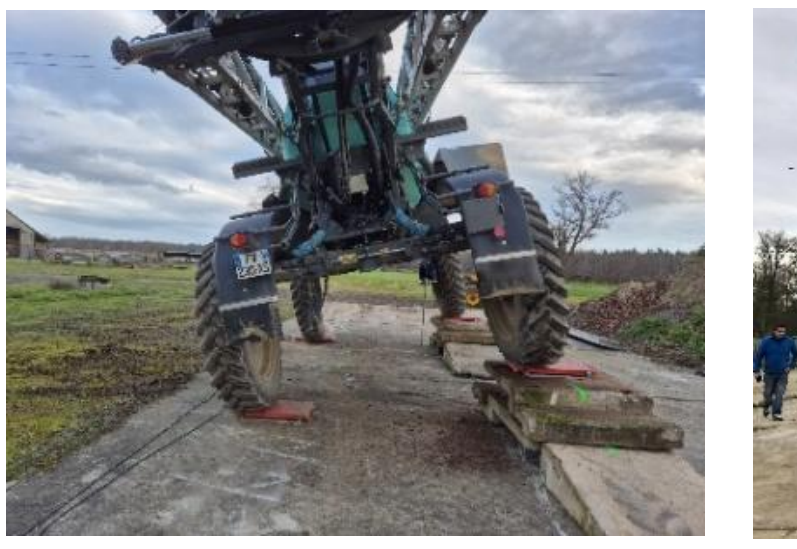
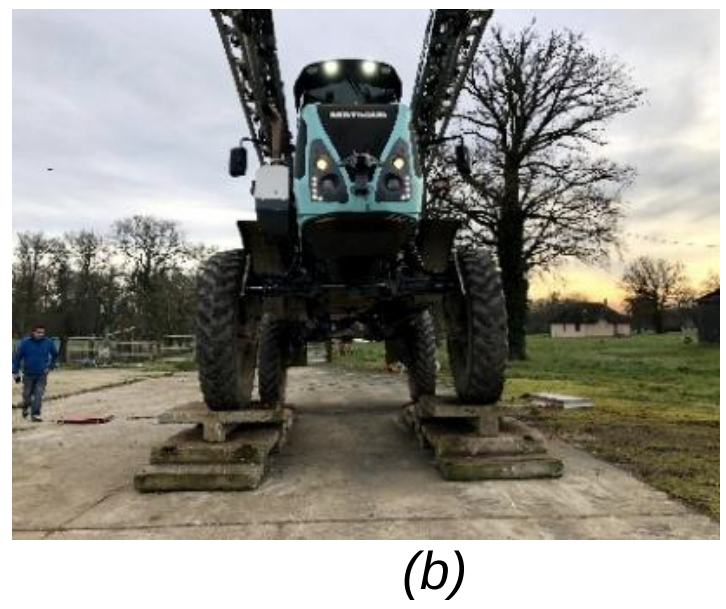
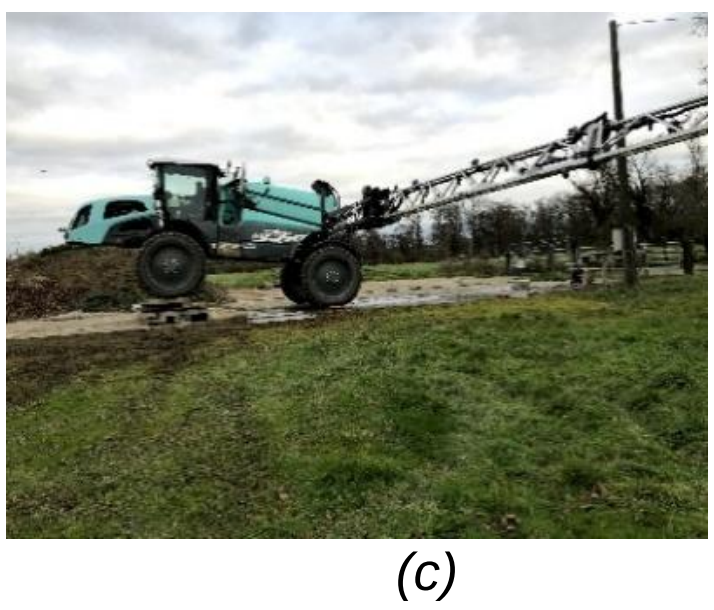

*(a) (b) (c)*

*Figure 7. Measurement of wheel-ground force distribution with the vehicle tilted laterally (a), and longitudinally with the two ramp configurations (b) closed, and (c) open*

*2.4.2. Dynamic tests*

Dynamic tests aim to acquire data during the navigation of the machinery. The parameters that vary in these tests include navigation speed (low or high), implement configuration (folded or unfolded), position of the unfolded implement (low, high or in a varying position), load or tank filling (empty, 50% full or 100% full), trajectory (straight line or turns) and ground conditions (slopes, inclination, asphalt road, path with natural soil or grassy field).

Both on-road and off-road navigation are carried out. On-road navigation is carried out on roads and paths belonging to Inrae Montoldre site, mainly on the AgroTechnoPôle test track (Figure 8). This test track has a flat, asphalt surface in the shape of an L, measuring 150 m by 100 m. It allows 90° turns and U-turns, as well as high-speed navigation. Low and high speeds adopted during tests on this track are 13 km/h and 25 km/h respectively.

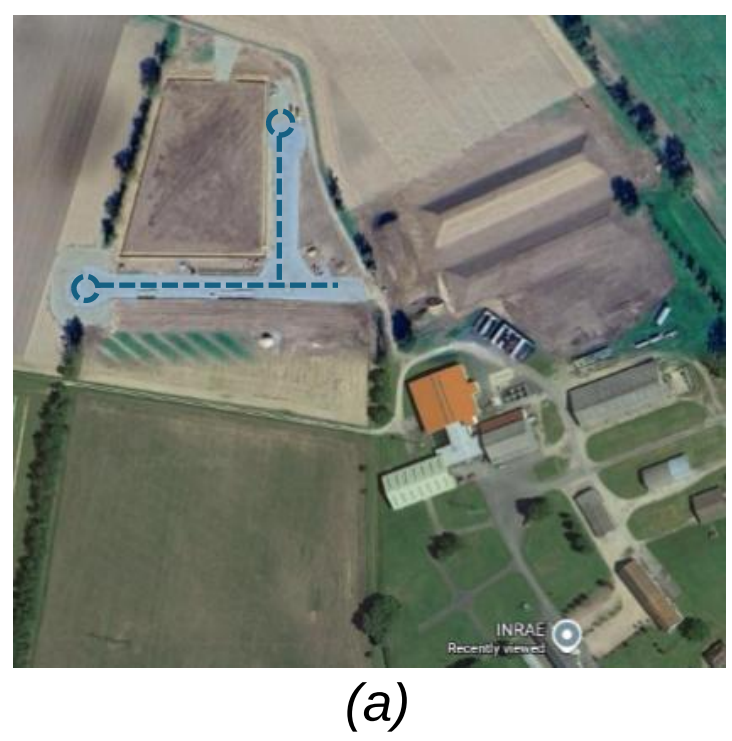
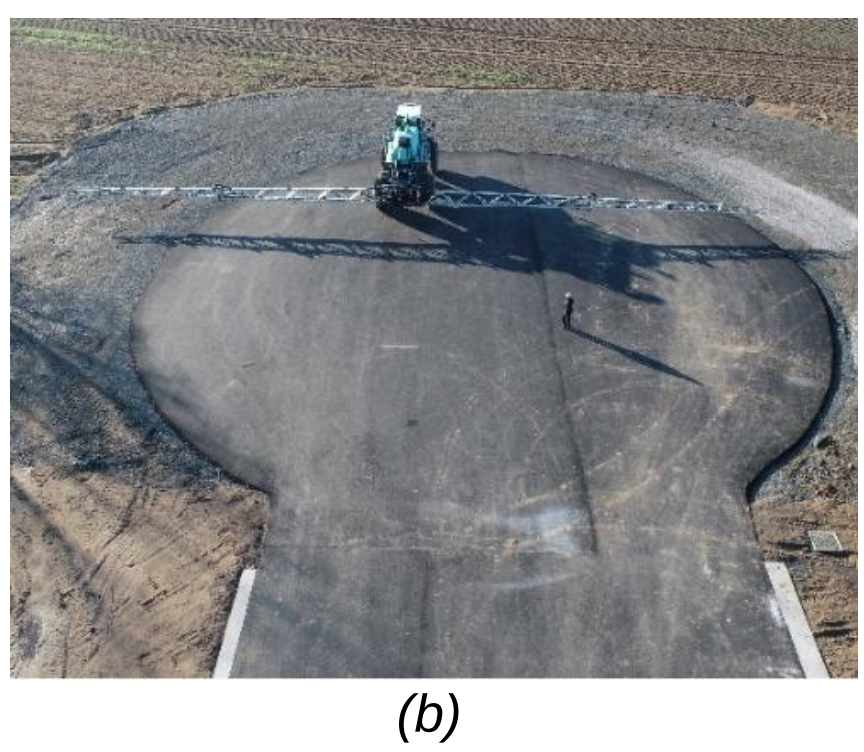

*(a) (b)*

*Figure 8. (a) An aerial view of the area and test track used for dynamic on-road navigation tests (extracted from Google Maps). (b) A view of the vehicle during on-road testing*

Off-road navigation is carried out on sloping agricultural land belonging to INRAE-Montoldre (Figure 9), which allows U-turns and manoeuvres on unevent areas with unfolded implements. The low and high speeds adopted during tests on this terrain are 6 km/h and 12 km/h respectively. Climatic conditions and the high level of soil humidity prevent the completion of all planned trial repetitions; however, sensor data is obtained during all 50 tests of vehicle navigation. These recordings include online load transfer estimation.

[1] https://www.agrotechnopole.fr/

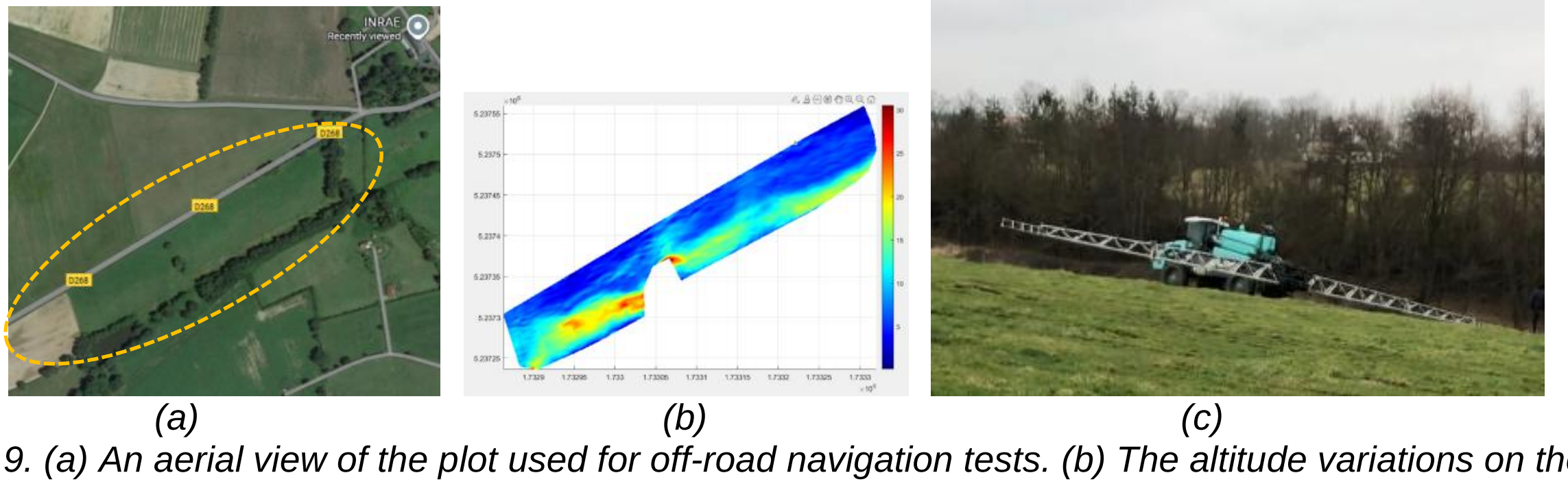

(a) (b) (c)

*Figure 9. (a) An aerial view of the plot used for off-road navigation tests. (b) The altitude variations on the plot, (c) A view of the vehicle during an off-road test*

## 3. Results and Discussion

The results of the work carried out are the data obtained, which are mainly used to feed the stability estimation algorithms.

### 3.1. Vehicle parameters

Six sets of parameters were obtained for the vehicle, corresponding to three levels of tank filling and two implement configurations.

As the tests to determine the position of the centre of gravity (particularly the vertical component, $H$) produced disparate values, only some of the test repetitions were considered. Based on these tests, mathematical relationships were obtained that determine the position of the vehicle's centre of gravity (its height $H$ and its longitudinal position, given by the half rear wheelbase $L_r$) as a function of the implement's position, as represented by the implement cylinder output ($x$) (see Figure 10).

The obtained parameters do not include the vehicle's moments of inertia or the roll and pitch damping coefficients. The former are obtained from the vehicle's CAD model, and the latter from expert feedback.

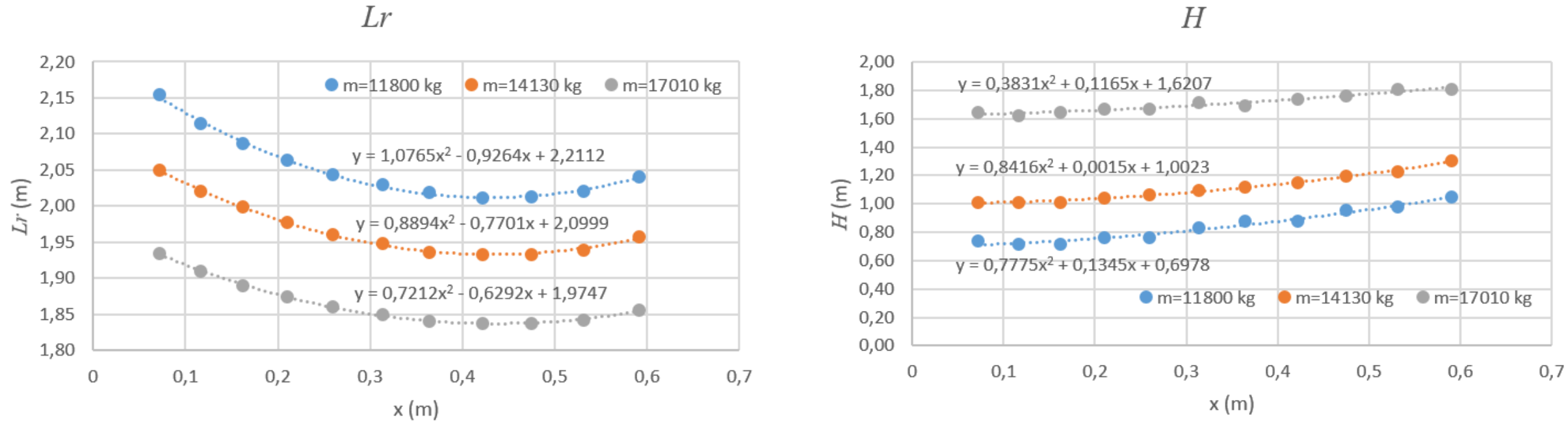

*Figure 10. The position of the vehicle's centre of gravity for three different vehicle masses, expressed as a function of the implement cylinder output (x). The centre of gravity position is given by the rear half wheelbase (Lr) and its height from the ground (H).*

### 3.2. Data related to the movement of the vehicle

Following the dynamic tests carried out in the field at the INRAE site in Montoldre, interesting measurement data was recorded. Figure 11 show an example of data acquired during on-road and off-road navigation tests. In both examples, the implement is in an unfolded configuration. In the first example (Figure 11(a)), the implement moves up and down as the vehicle moves. The vehicle has a full tank and follows straight-line trajectories involving right-angled turns and U-turns. In the second example (Figure 11 (b)), the vehicle moves off-road on slippery, sloped ground with a 50% tank capacity. The implement remains in a fixed position during vehicle movement.

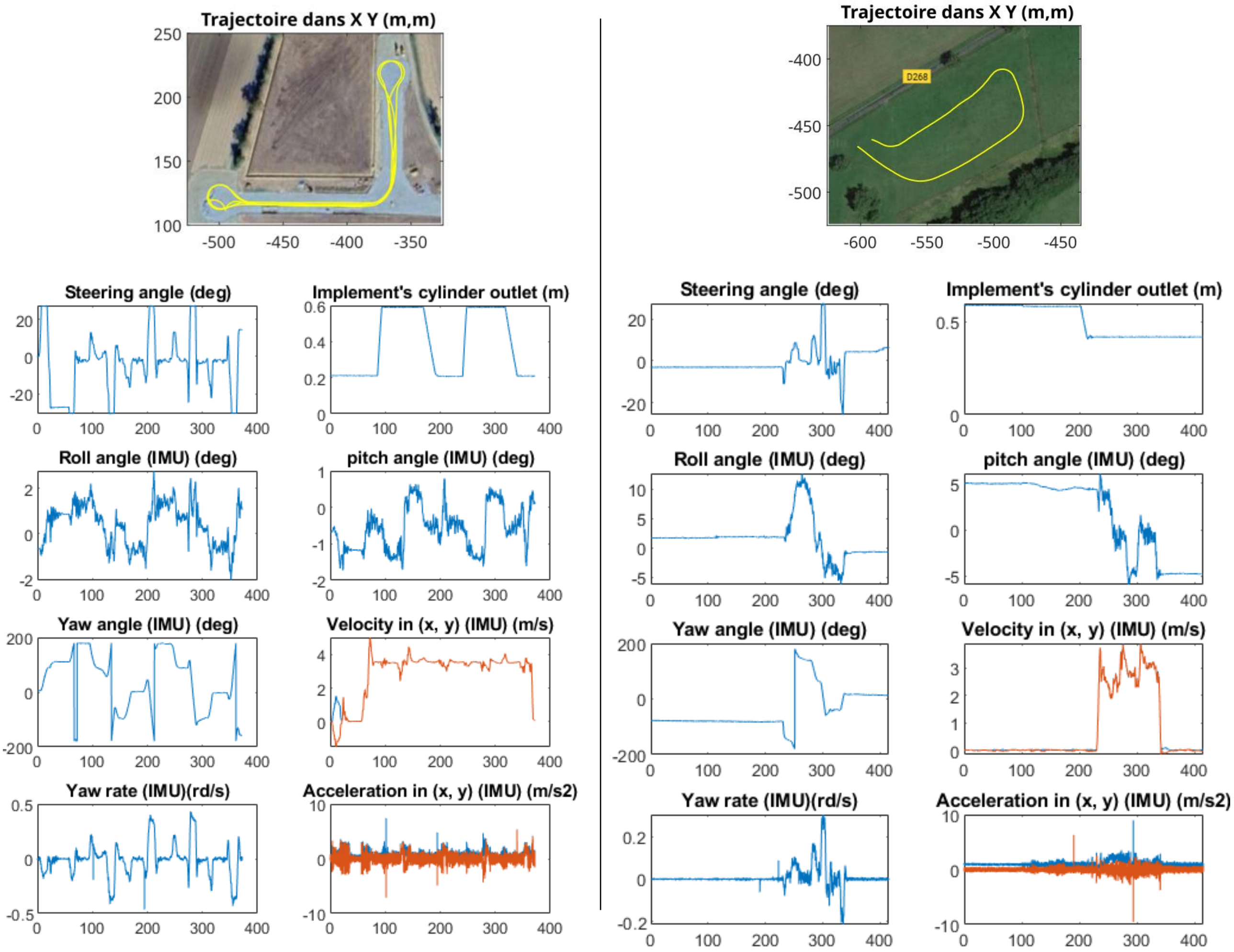

Trajectoire dans X Y (m,m)

Steering angle (deg)

Implement's cylinder outlet (m)

Roll angle (IMU) (deg)

pitch angle (IMU) (deg)

Yaw angle (IMU) (deg)

Velocity in (x, y) (IMU) (m/s)

Yaw rate (IMU)(rd/s)

Acceleration in (x, y) (IMU) (m/s2)

(a)

Trajectoire dans X Y (m,m)

Steering angle (deg)

Implement's cylinder outlet (m)

Roll angle (IMU) (deg)

pitch angle (IMU) (deg)

Yaw angle (IMU) (deg)

Velocity in (x, y) (IMU) (m/s)

Yaw rate (IMU)(rd/s)

Acceleration in (x, y) (IMU) (m/s2)

(b)

*Figure 11 Examples of data related to the movement of the vehicle acquired during (a) an on-road test and (b) an off-road test*

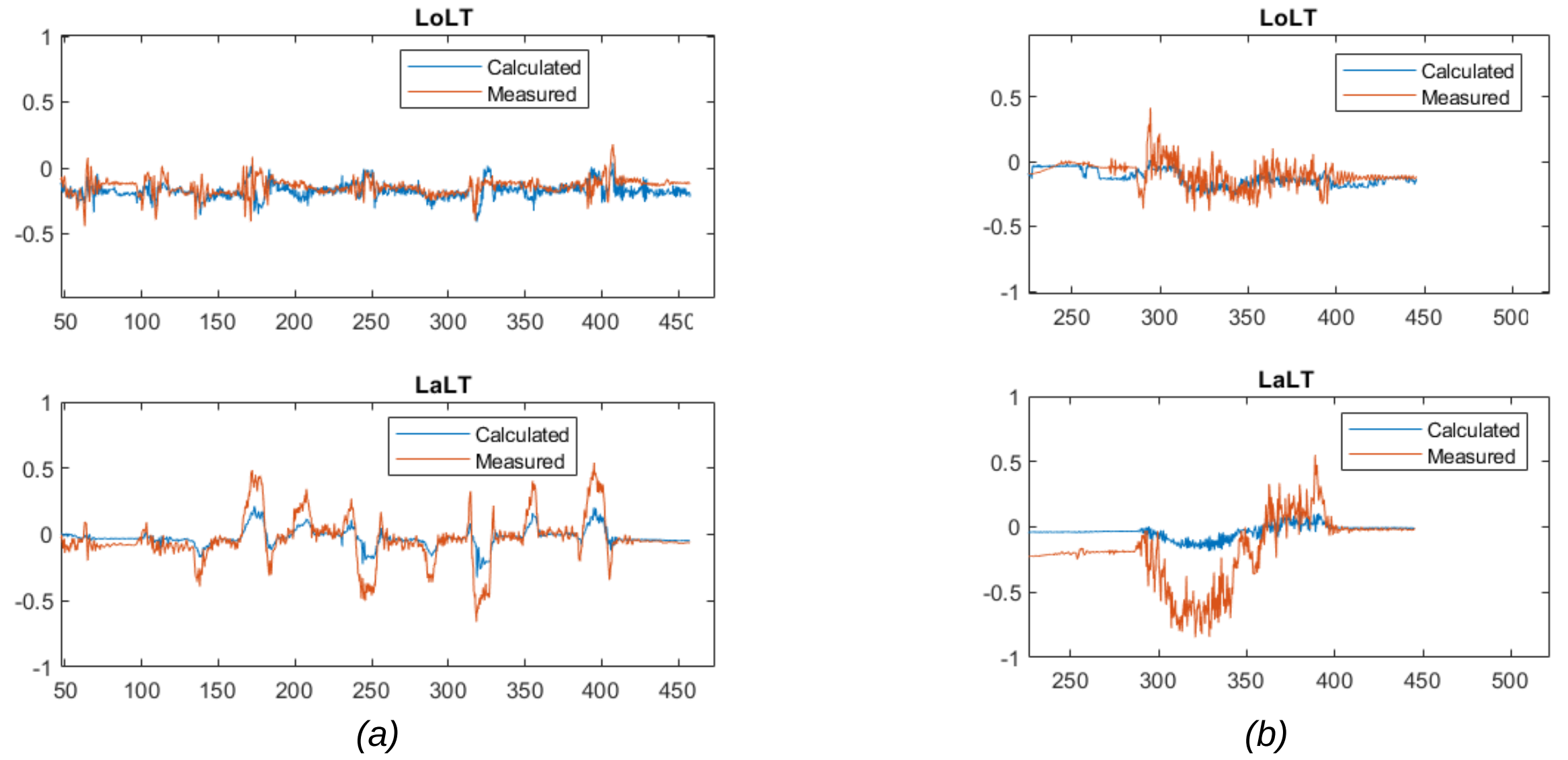

(a) (b)

*Figure 12. Examples of online load transfer, calculated using the algorithms (blue curves) and deduced from the strain gauge measurements (red curves), obtained during (a) an on-road test and (b) an off-road test (related to the cases described in Section 3.2).*

*3.3. Estimated and measured online load transfer*

The initial results of comparing the load transfers calculated by the algorithms with those obtained from strain gauge measurements demonstrate the algorithms' ability to capture variations in load transfer. However, the numerical values of the load transfers compared for each test show an offset (Figure 12). To address this, the next stage of this work will involve conducting a sensitivity study on the estimation algorithms to determine the impact of imprecise vehicle parameters on the estimation process. Additionally, some unused data will be employed to enhance the calibration model, providing wheel-ground forces from strain gauge measurements.

## Conclusions

This paper presents the experimental part of the work to test and evaluate the algorithms for estimating the rollover risk of machines with a mobile implement. After a presentation of mathematical models used in these algorithms, it describes the instrumentation choices made for a self-propelled sprayer with a mobile ramp. The instrumentation on the machine provided an instantaneous measurement of the wheel/ground force distribution, and thus the value of the 'measured' load transfer. This instrumentation also allowed some variables to be measured and processed by an on-board control unit in order to obtain, at each time step, the parameters of the machine and its evolution conditions, and to estimate the load transfer. The results of the comparison of the two load transfers ("measured" and "estimated") are presented and discussed.

## Acknowledgements

The authors thank Gama Technologies (Berthoud) for supplying the test vehicle, and the Auvergne-Rhône-Alpes region, the French government (FNADT fund), and the Allier department for supporting the AgroTechnoPôle platform, notably its off-road test tracks and operations.